\documentstyle[prb,aps]{revtex}
\input epsf
\begin{document}

\preprint{}
\title{Quantum wires from coupled InAs/GaAs strained quantum dots
}
\author{Craig Pryor\cite{email}}
\address{
Department of Solid State Physics \\
Box 118, Lund University\\
S-221 00 Lund, Sweden
}
\maketitle
\begin{abstract}
The electronic structure of an infinite 1D array of vertically coupled 
InAs/GaAs strained quantum dots is calculated
using an  eight-band strain-dependent $\bf k\cdot p$ Hamiltonian.
The coupled dots form a unique quantum wire structure in
which the miniband widths and effective masses
are controlled by the distance between the islands, $d$. 
The miniband structure is calculated as a function of $d$, and it is shown
that for $d>4~\rm nm$ the miniband is narrower than the optical phonon energy, 
while the gap between the first and second minibands
is greater than the optical phonon energy.
This leads to decreased optical phonon scattering,
providing improved  quantum wire behavior at high temperatures. 
These miniband properties are also ideal for  Bloch oscillation.
\end{abstract}


Semiconductor heterostructures have made possible the construction
of low-dimensional electronic systems. 
By providing confining barriers in one, two, or
three dimensions one obtains a quantum well, wire, or dot respectively.
It is also possible to increase the dimensionality by coupling a series of
low dimensional structures. Here we consider a quantum wire formed
by an infinite 1D array of quantum dots. Since the properties of the wire
are sensitive to the tunnelling between dots, they may
be tuned in a manner that is not possible with other quantum wire structures.
Such a structure has been proposed for designing heterostructures with
reduced optical phonon coupling.\cite{sakaki_1}

Periodic arrays of quantum dots have been constructed using stacks of
Stranski-Krastanov  islands. 
\cite{experiments_2,models_3}
In Stranski-Krastanov 
growth a lattice mismatched semiconductor is epitaxially
deposited on a substrate material. Due to the mismatch,
the deposited material beads up into nm-scale islands, that are 
subsequently covered with barrier material.
If only a small amount of barrier material is deposited over the island, 
followed immediately by
another island deposition, the new islands form directly over the previously
deposited islands. By repeating the procedure a series of self-aligned
quantum dots is obtained. Since the periodicity is determined by the barrier
deposition, it may be accurately controlled.

Photoluminescence experiments on InAs/GaAs island stacks
are in rough agreement with estimates using simple 1D
models.\cite{models_3}
These estimates were based on repeating the single-band potential from
an isolated island, neglecting the strain effects
from neighboring islands and band mixing. Also, the full 3D structure
were not taken into account. Here we compute the miniband structure
employing  the full 3D structure, realistic strain, and band mixing.

\vbox{
\begin{figure}
\epsfxsize=8.75cm
\epsfbox{ 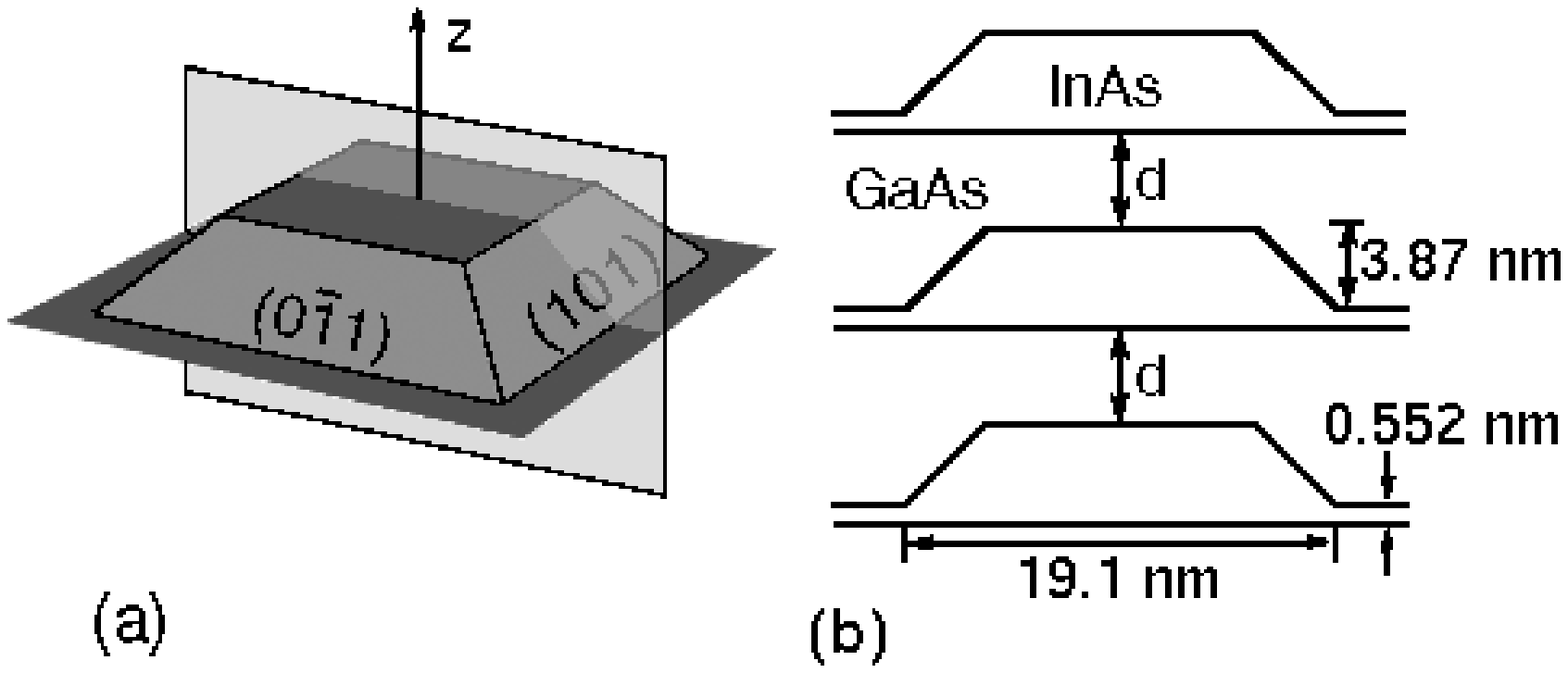 }
\caption{ Island geometry. (a) a single InAs island (b) a cross section
of the stack through the plane indicated in (a).
\label{fig1}
}
\end{figure}
}

The specific system considered consists of stacked InAs islands 
embedded in GaAs, as shown in Fig. 1.  Each island is a truncated
square-based pyramid with $(101)$-type faces. The islands
are  $19.1~\rm nm$ wide, $3.87~\rm nm$ high, and include
a $0.552 ~\rm nm$ thick wetting layer. 
The size and shape are in agreement with transmission electron micrographs,
though there is considerable uncertainty in the island geometry.
The precise numbers used here 
come about because the calculation is done on a grid
commensurate with the wetting layer thickness. Thus, all dimensions are 
multiples of the grid spacing, which is fixed by the choice of wetting
layer thickness.
The islands are assumed to be of fixed size and shape, and only the
distance $d$ between them is allowed to vary. 
We consider  the range
$d = 0.552 ~{\rm nm}\approx 2 ~\rm monolayers$ to
$d = 4.97  ~{\rm nm}\approx 18 ~\rm monolayers$.
The wetting layer is potentially  problematic if it is thick enough
for the electron wave function to significantly
penetrate into the wetting layer.  In order to consider
the worst case scenario, we assume the wetting layer is 
$0.552 ~\rm nm$ thick, which corresponds to  two  monolayers of InAs
biaxially strained to match the GaAs substrate.
Most measurements have found the wetting layer to be less than two monolayers
thick.

The calculational method has been described in detail elsewhere.\cite{pryor_4}
Continuum elastic theory is used to compute the strain by
discretizing the system on a cubic grid and 
numerically minimizing the strain energy. 
The result is used as input to a strain-dependent eight-band 
envelope-approximation Hamiltonian. 
The strain induced piezoelectric charge is also included as an additional
potential.
The Hamiltonian is discretized on the same grid, and its eigenvalues are
found by Lanczos diagonalization.
An eight-band model is used because InAs has a narrow band gap, 
making band mixing significant.
The narrow gap and large strain  result in a near doubling of the electron 
effective mass within the island\cite{cusack_5,pryor_6} which would
be neglected in a naive single band approximation.
All material parameters are set to the values used in reference 4.

\vbox{
\begin{figure}
 \hbox{
  \epsfxsize=8.75cm
  \epsfbox{ 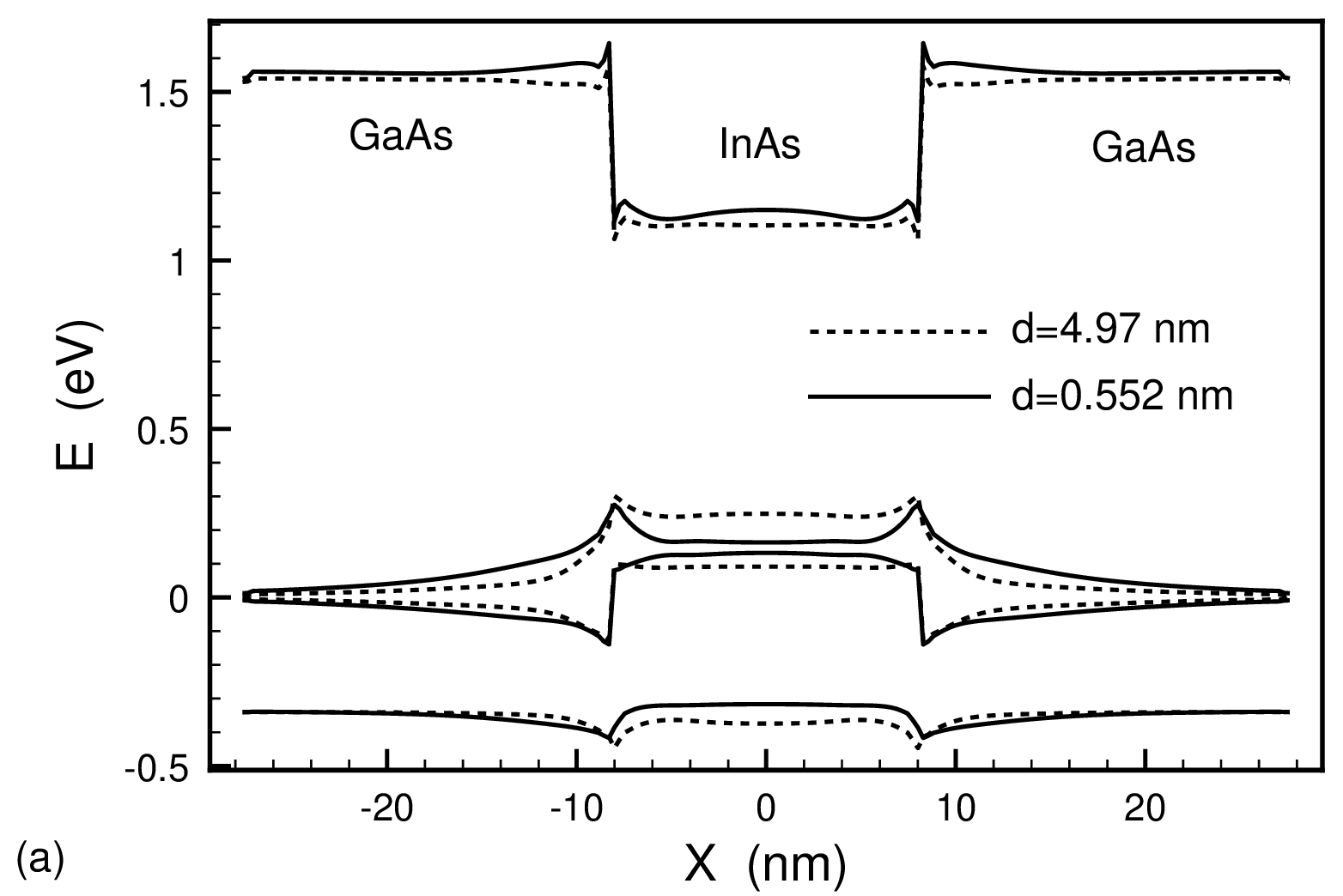 }
  \epsfxsize=8.75cm
  \epsfbox{ 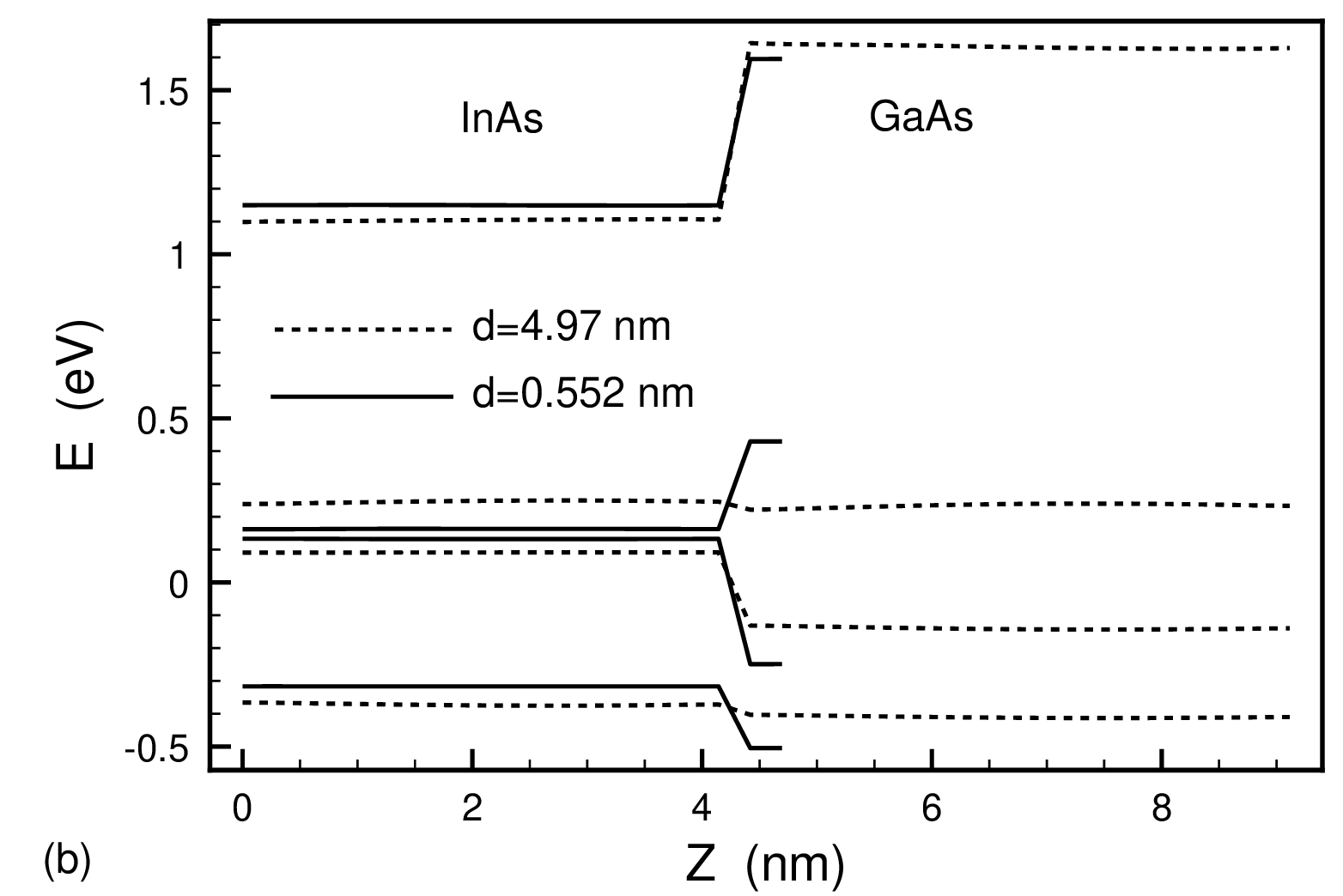 }
 }
\caption{ Band structures for $d=  0.552 ~{\rm nm}$ and $d=  4.97 ~{\rm nm}$.
(a) along the 100 direction, through the center of an island 
(b) one period along the 001 direction along the axis of the stack.
\label{fig2}
}
\end{figure}
}

Fig. 2 shows the band structures for 
$d=  0.552 ~{\rm nm}$ and $d=  4.97 ~{\rm nm}$. These were obtained
by diagonalizing the Hamiltonian, with $\vec k = 0$ and using the local
value of the strain.
Because the strain extends into the barrier material, the strain
within one island is affected by neighboring islands as well.
The band diagrams along the direction perpendicular to the wire show
good 1D confinement for both electrons and holes.
The electrons are confined by a barrier of approximately  $350 ~\rm meV$
for $d= 0.552 ~\rm nm$, and $400 ~\rm meV$ for $d= 4.97 ~\rm nm$.
The hole confinement is more sensitive to $d$, varying from 
$180 ~\rm meV$ to  $250~\rm meV$ as measured at the center of the island.
The band structure along the axis of the wire is shown in Fig 2b.
For electrons the on-axis potential varies by about $50 ~\rm meV$
over the range of $d$.
Changing $d$ primarily alters the electron
barrier thickness with little change in the shape of the potential within
the island.
The valence band is more complex.
For small separations the GaAs between the islands is highly
strained,  resulting in a valence band edge that is {\it higher} in the GaAs
than in the InAs. This type II behavior is only seen in the potential along the
wire axis, and the transverse potential is always confining.
For $d>4.5 ~\rm nm$ the barrier strain is sufficiently small
that the deepest confining hole potential is in the InAs island.
The barrier between islands is, however,  only about $10 ~\rm meV$ for
$d=  4.97 ~{\rm nm}$.

\vbox{
\begin{figure}{
 \hbox{
  \epsfxsize=8.75cm
  \epsfbox{ 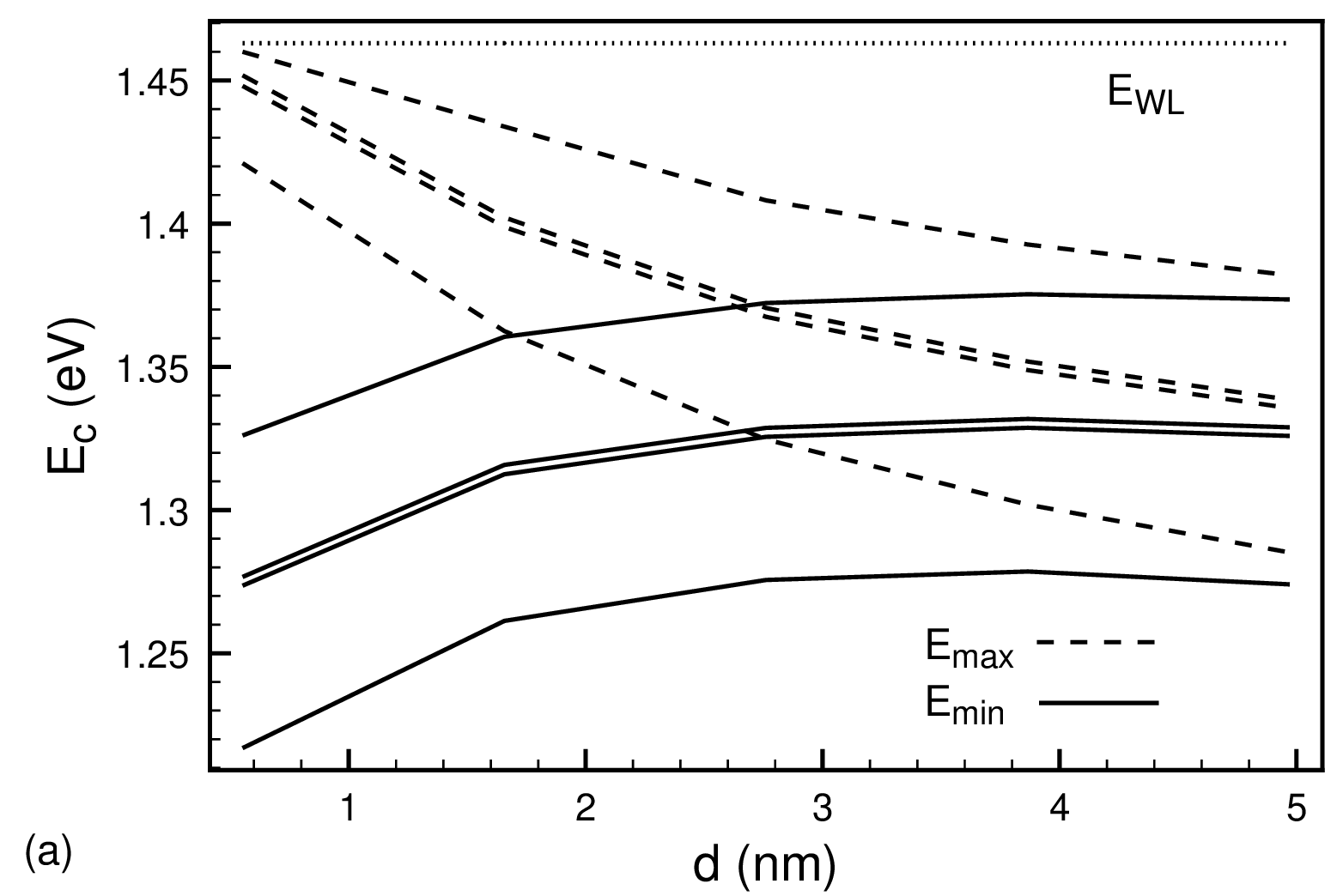 }
  \epsfxsize=8.75cm
  \epsfbox{ 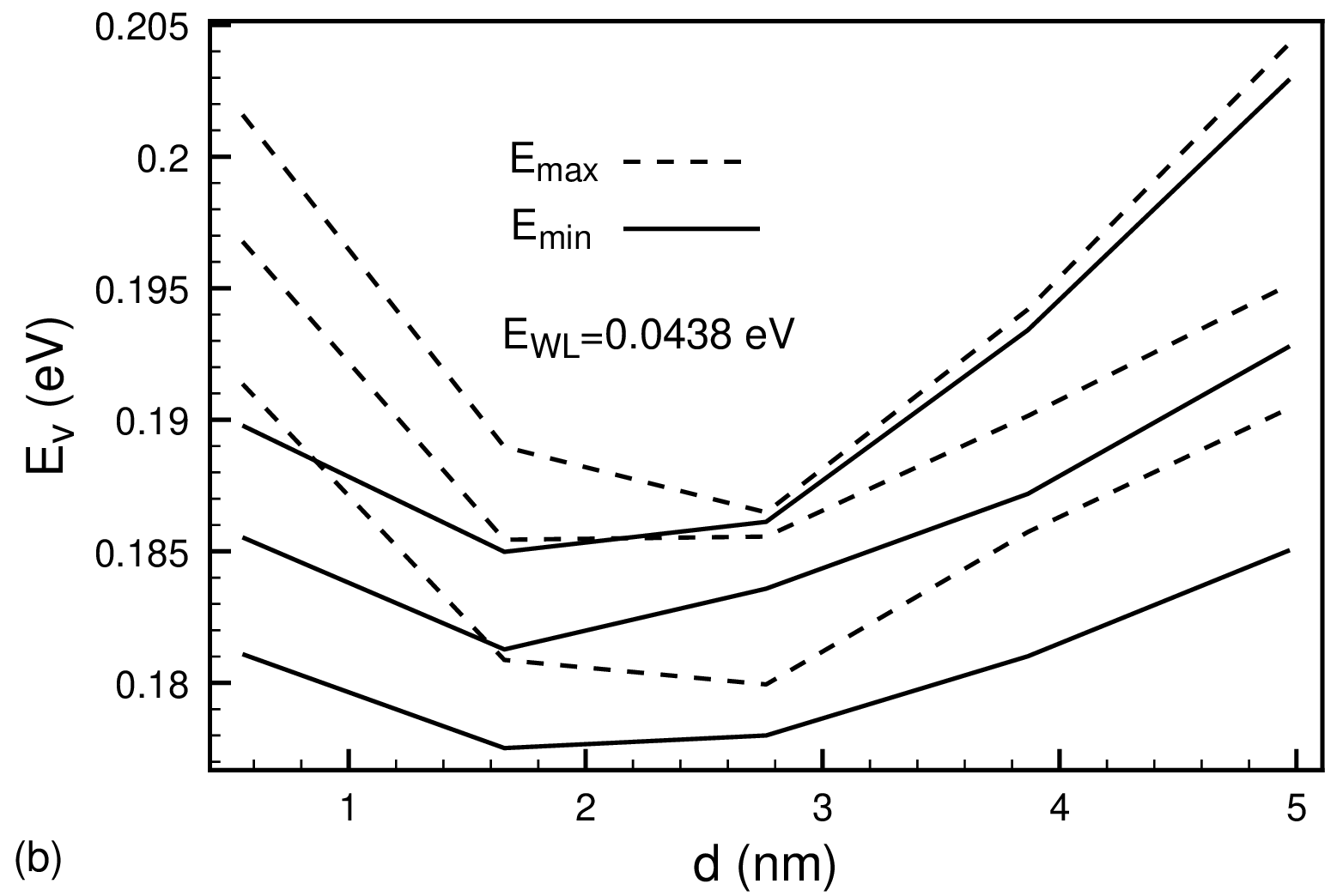 }
 }

\caption{ Miniband widths as a function of island spacing. Solid lines
  are the minimum energies for each miniband, dashed lines are the
  maxima. All energies are with respect to
  the valence band edge in GaAs infinitely far from the islands 
  (i.e. unstrained).
  (a) The lowest four conduction minibands.  
  The dotted line is the energy for the single bound state in the wetting
  layer.
  (b) The highest three valence minibands. 
  The wetting layer ground state energy, $E_{WL}$, is below the
  range of the graph.
\label{fig3}
}
}
\end{figure}
}

The minimum and maximum miniband energies 
are plotted in Fig. 3 as a function of island separation.
The conduction band states are the most interesting since they have the 
largest subband spacings.
The  minima and maxima are at $k_z=0$ and $k_z=\pm \pi / L_z$
respectively,  where $k_z$ is the momentum along the wire, and
  $L_z$ is the period.
For $d>2.75 ~\rm nm$ the first miniband separates from the other minibands,
although the second and third minibands still overlap since
they are nearly degenerate. (The small splitting is due to 
the piezoelectric charge that breaks the $C_4$ symmetry of the square island
down to $C_2$.\cite{pryor_4,grundmann_7} )
The most interesting regime is  $d>4 ~\rm nm$, where
the gap between the first and second minibands is 
greater than $30 ~\rm meV$, and  the width of the first miniband is less than
$20 ~\rm meV$.
Since the optical phonon energy is approximately $30 ~\rm meV$, both
interband and intraband transitions will be suppressed;
 an electron in the lowest 
miniband has no final states available one optical phonon energy away.
Since optical phonon scattering is dominant at high temperatures, the
structure described here should maintain quantum wire behavior to higher
temperatures than ordinary wires. 

\vbox{
\begin{figure}
\epsfxsize=8.75cm
\epsfbox{ 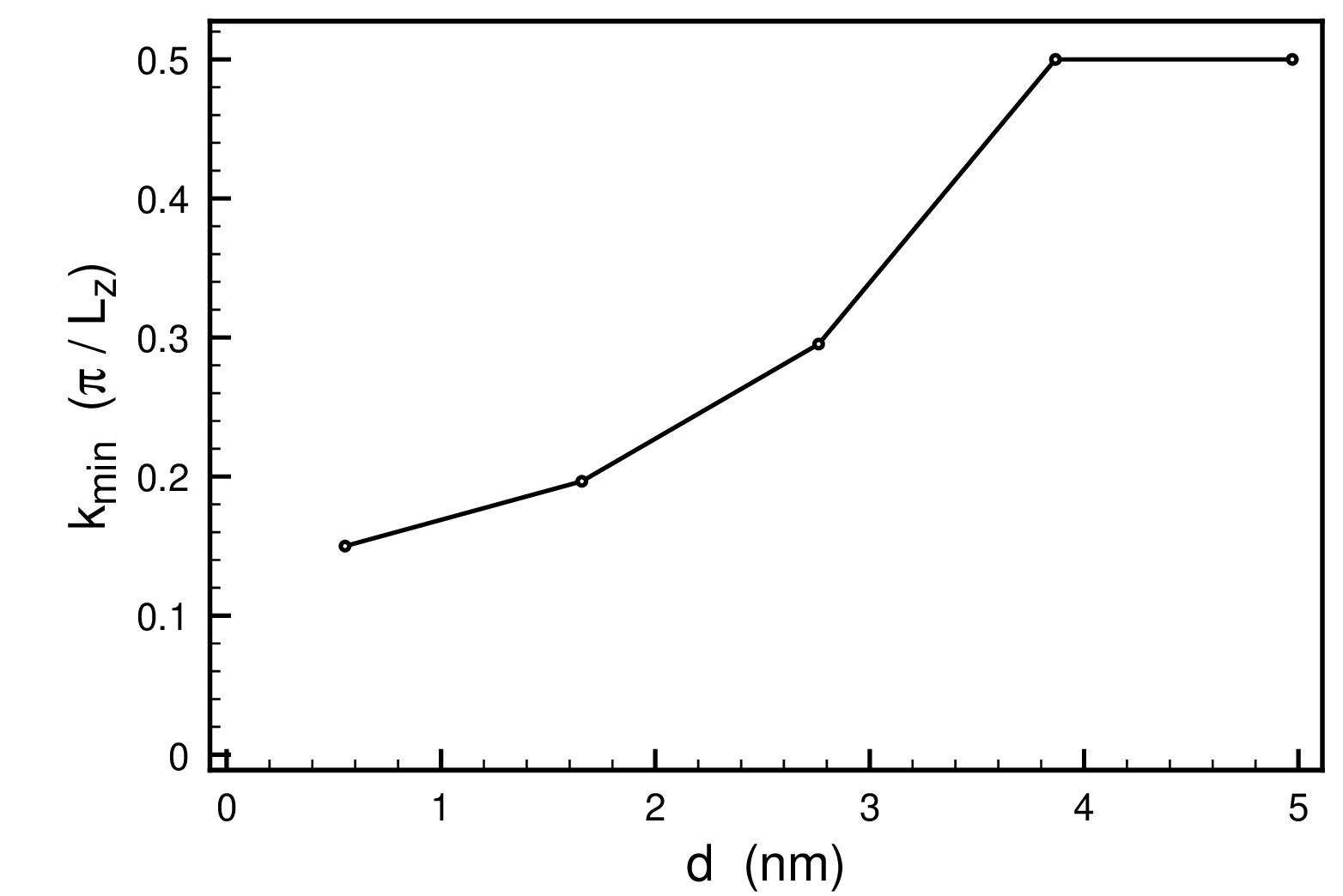 }
\caption{ Location of valence miniband maximum, expressed in units of the
minizone width.
\label{fig4}
}
\end{figure}
}

In contrast to the conduction band,
the valence band structure is considerably more complex (Fig. 3b).
The miniband spacings range from  $5 ~\rm meV$  to $10 ~\rm meV$
while the miniband widths vary from $10 ~\rm meV$  to less than $1~ \rm meV$.
For $d>3 ~\rm nm$ the three highest valence minibands are non-overlapping,
while for $d<1.6 ~\rm nm$ there are no gaps. At large values of $d$ the
miniband widths are sufficiently small to suppress optical phonon scattering,
however the gaps between minibands are too small to suppress interminiband
scattering.  
The valence miniband minima all occur at $k_z=\pi / L_z$ and the maxima
occur away from zone center. Such indirect gaps are
also seen in standard quantum wires.\cite{sercel_8}  
Fig. 4 shows
the value of $k_z$ for which $E_v(k_z)$ is a maximum, as a function of $d$.
For $d >  4 ~\rm nm$ the miniband maximum occurs at the minizone edge, while
for smaller separations the maximum occurs in the middle of the zone.

\vbox{
\begin{figure}{
\epsfxsize=8.75cm
\epsfbox{ 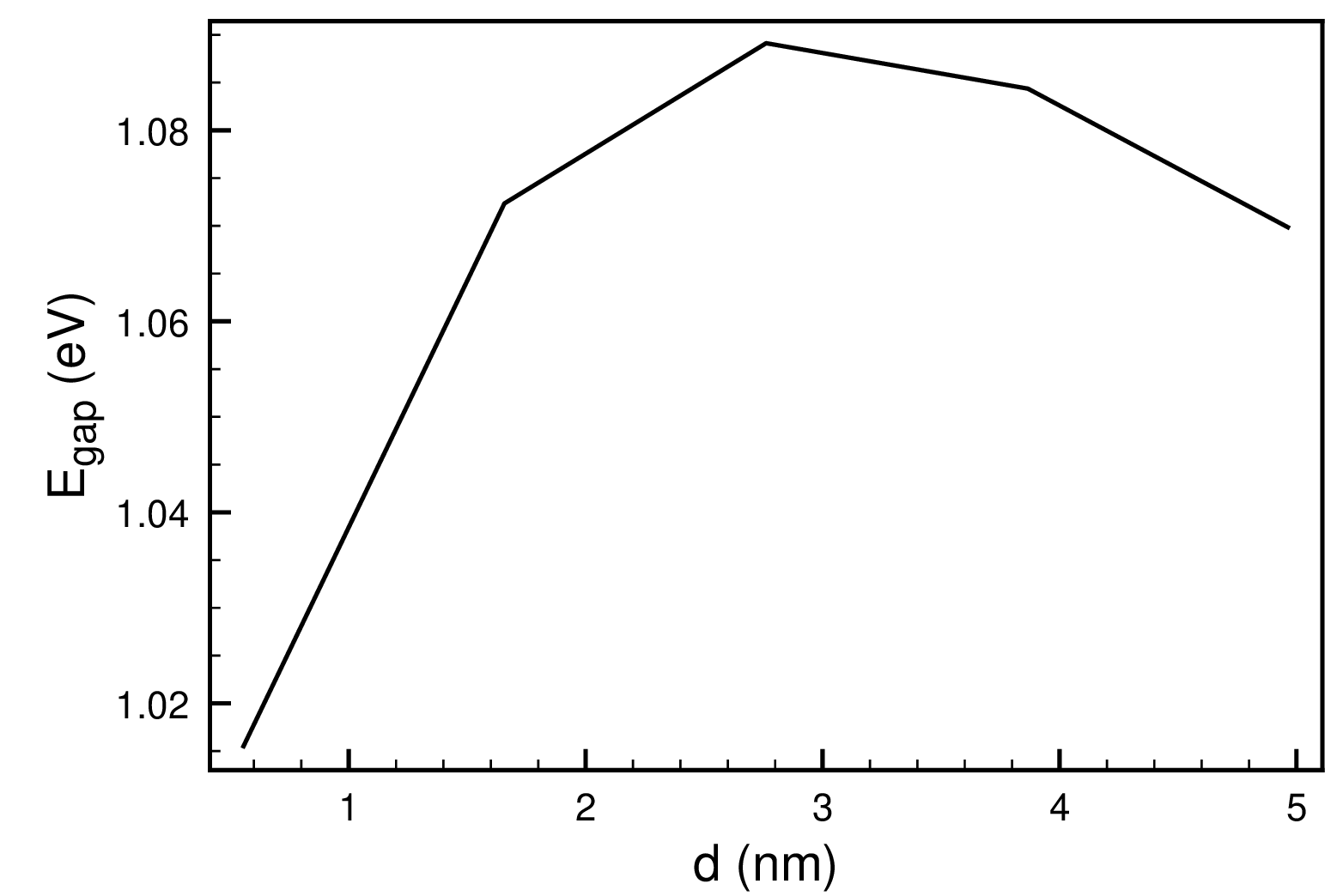 }
\caption{ Miniband gap as a function of island seperation.
\label{fig5}
}
}
\end{figure}
}

While the miniband widths determine the quantum wire properties of the
structure, they are not directly measurable. A  
quantity such as the miniband gap is useful for characterizing the structure
since it may be measured using photoluminescence.
Fig. 5 shows the miniband gap as a function of island separation. For
simple coupled quantum dots one expects a minimum energy at $d=0$,
monotonically  increasing with $d$. However, due to the fact that $d$ 
affects the strain, the gap turns over, giving a maximum gap at 
$d\approx 3 ~\rm nm$. Photoluminescence measurements  give
$1.08 ~{\rm eV} < E_{gap}<1.15 ~{\rm eV}$
for stacks containing up to 10 islands and 
with $d=1.5 ~\rm nm$.\cite{models_3} The
gap shown in Fig. 5 is somewhat smaller ( $E_{gap}\approx 1.06 ~\rm eV$),
however this slight difference is not significant given the uncertainties in
the island size.

\vbox{
\begin{figure}
\epsfxsize=8.75cm
\epsfbox{ 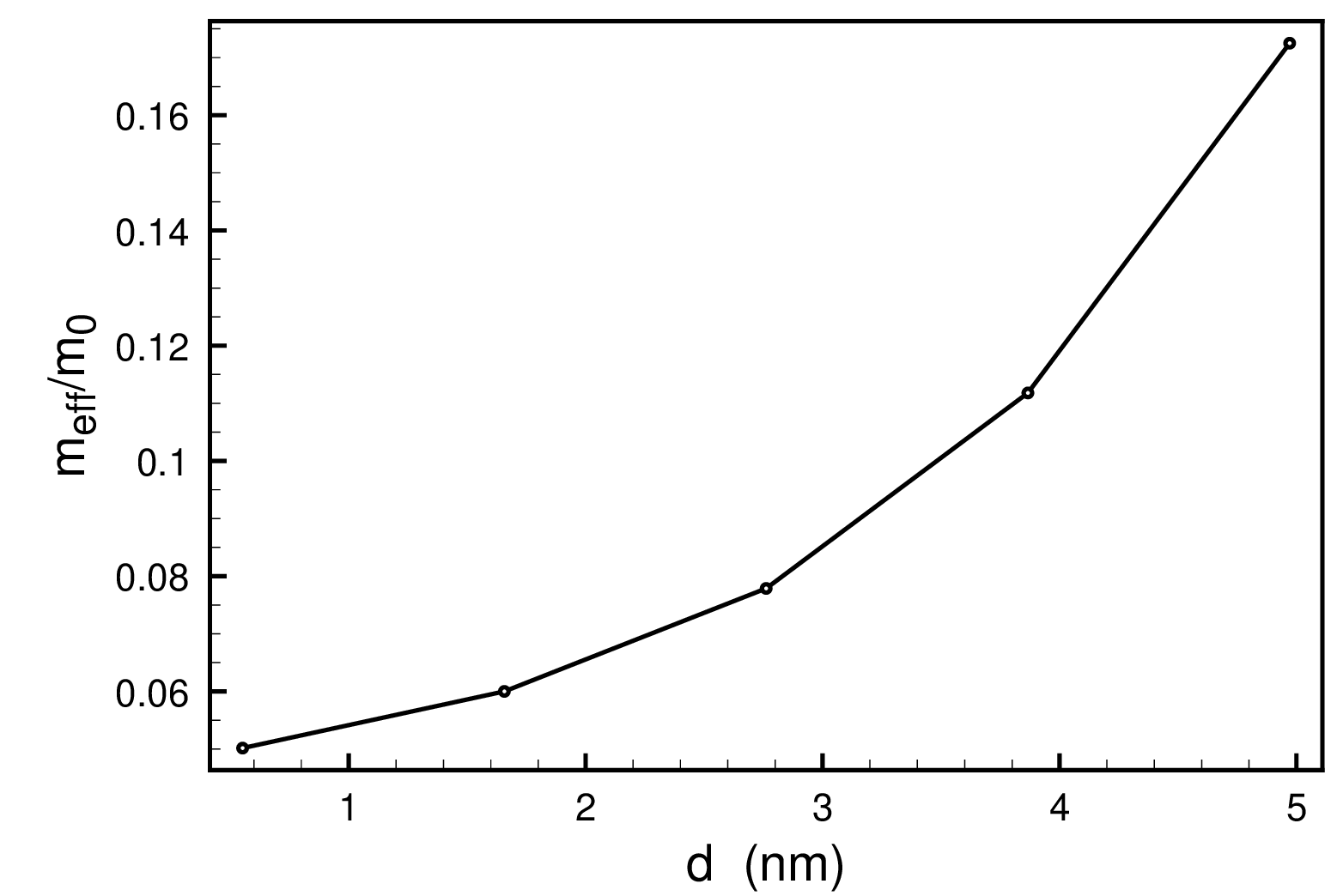 }
\caption{Conduction band effective mass a function of island spacing $d$.
\label{fig6}
}
\end{figure}
}

The miniband structure may also be parameterized by the
effective mass, which  was calculated for the first conduction miniband
by fitting a  parabola to $E(k_z)$ around $k_z=0$  (Fig 6).
For contiguous islands $m_{eff}\approx 0.05 m_0$, which is comparable
to the value calculated within a  single InAs islands.\cite{cusack_5,pryor_6}
With increasing barrier thickness the effective mass increases
quickly, reflecting the exponentially suppressed tunnelling probability.
This ability to tune the effective mass allows a degree of band engineering
not available with standard wire structures.

The quantum wire structure described above is particularly well suited
to producing Bloch oscillations.\cite{blochRev_9} 
Bloch oscillations have been measured
in superlattices, however, due to dephasing effects the oscillations
are heavily damped and only a few periods
are observed. The wire structure presented here lacks the transverse
excitations present in superlattice structures, eliminating this
source of damping. At low carrier densities the damping has been attributed
to optical phonon scattering\cite{rossi_10}. Thus island stacks with
$d > 4 ~\rm nm$ should make improved Bloch oscillators. 
Damping is also produced by imperfections in 
the periodicity\cite{reynolds_11}, which of course  island stacks also 
suffer from  due to growth variations. 

In conclusion, we have seen that quantum wires with non-overlapping 
 minibands may  be obtained
from vertically coupled strained InAs/GaAs quantum dots.
For  an island spacing
$d > 4 ~\rm nm$ the lowest electron miniband width is less than the
optical phonon energy, and the gap to the second miniband is greater than the
optical phonon energy. This results in quantum wires with
 decreased optical phonon coupling.
The same range, $d > 4 ~\rm nm$, gives a miniband structure favorable
for generating Bloch oscillations.

I wish to thank Mark Miller for stimulating discussions.

\end{document}